# COVID-19 Policy Impact Evaluation: A guide to common design issues

Noah A Haber, Emma Clarke-Deelder, Joshua A Salomon,  Avi Feller, Elizabeth A Stuart

Noah A Haber, ScD*
noahhaber@stanford.edu
Meta Research Innovation Center at Stanford University
Stanford University
1265 Welch Rd
Palo Alto, CA 94305
(650) 497-0811

Emma Clarke-Deelder, MPhil
Department of Global Health & Population
Harvard T. H. Chan School of Public Health
665 Huntington Avenue
Building 1, room 1104
Boston, Massachusetts 02115

Joshua A Salomon, PhD
Department of Medicine
Center for Health Policy and Center for Primary Care and Outcomes Research
Stanford University School of Medicine
Encina Commons, Room 118
615 Crothers Way
Stanford, CA 94305-6019

Avi Feller, PhD
Goldman School of Public Policy
University of California, Berkeley
2607 Hearst Avenue
Room 309
Berkeley, CA 94720

Elizabeth A Stuart, PhD
Department of Mental Health
Johns Hopkins Bloomberg School of Public Health
624 N. Broadway
Hampton House 839
Baltimore, MD 21205

* corresponding author

## Abstract

Policy responses to COVID-19, particularly those related to non-pharmaceutical interventions, are unprecedented in scale and scope. Epidemiologists are more involved in policy decisions and evidence generation than ever before. However, policy impact evaluations always require a complex combination of circumstance, study design, data, statistics, and analysis. Beyond the issues that are faced for any policy, evaluation of COVID-19 policies is complicated by additional challenges related to infectious disease dynamics and lags, lack of direct observation of key outcomes, and a multiplicity of interventions occurring on an accelerated time scale. The methods needed for policy-level impact evaluation are not often used or taught in epidemiology, and differ in important ways that may not be obvious. The volume and speed, and methodological complications of policy evaluations can make it difficult for decision-makers and researchers to synthesize and evaluate strength of evidence in COVID-19 health policy papers.

In this paper, we (1) introduce the basic suite of policy impact evaluation designs for observational data, including cross-sectional analyses, pre/post, interrupted time-series, and difference-in-differences analysis, (2) demonstrate key ways in which the requirements and assumptions underlying these designs are often violated in the context of COVID-19, and (3) provide decision-makers and reviewers a conceptual and graphical guide to identifying these key violations. The overall goal of this paper is to help epidemiologists, policy-makers, journal editors, journalists, researchers, and other research consumers understand and weigh the strengths and limitations of evidence that is essential to decision-making.

## Introduction

The COVID-19 pandemic has demanded urgent decision making in the face of substantial uncertainties. Policies to reduce transmission, including stay-at-home orders and other non-pharmaceutical interventions (NPIs), have wide-reaching consequences. Decision-making in the public interest requires evaluating and weighing the evidence on both intended and unintended consequences.[1–3] Epidemiologists are increasingly being asked to participate in policy recommendations, evidence evaluation, and evidence generation. Given the importance of these problems, there has been a proliferation of studies aiming to evaluate interventions implemented by different jurisdictions to inform future policymaking.[4] To be informative, however, such evaluations require a complex combination of circumstance, data, study design, analysis, and interpretation.

Estimating the causal impact of the policy with observational data is challenging because what would have happened in the absence of the policy change (the "counterfactual") is, by definition, unobserved. Randomized controlled trials (RCTs) of policies related to COVID-19 interventions may not always be practical or ethical.[5] However, there are many potential pitfalls and methodological design challenges relating to impact evaluation in general and COVID-19 policies in particular.

While many of the basic ideas behind longitudinal impact evaluation methods harken back to John Snow, they are relatively rarely taught and practiced in contemporary epidemiology,[6,7] particularly with regard to the intricacies of policy evaluation. As Caniglia and Murray note, in epidemiology, these methods have "fallen out of use given the field's shift to focus on questions of clinical relevance rather than population health relevance."[6] That is changing, with these types of methods increasingly being included in epidemiology curricula, and chapters being added to texts, such as Glymour and Swanson's recent chapter addition in the 4th edition of *Modern Epidemiology*.[8] These efforts are well-timed, as COVID-19 has highlighted the need for epidemiology to rapidly re-engage with policy evaluation methods.

This paper provides a graphical guide to policy impact evaluations for COVID-19, targeted to decision-makers, researchers and evidence curators. Our aim is to provide a coherent framework for conceptualizing and identifying common pitfalls in COVID-19 policy evaluation. Importantly, this should not be taken either as a comprehensive guide to policy evaluation more broadly or as guidance on performing analysis, which may be found elsewhere.[9–13] Rather, we review relevant study designs for policy evaluations — including pre/post, interrupted time series, and difference-in-difference approaches — and provide guidance and tools for identifying key issues with each type of study as they relate to NPIs and other COVID-19 policy interventions. Improving our ability to identify key pitfalls will enhance our ability to identify and produce valid and useful evidence for informing policymaking.

# Common policy evaluation designs and their pitfalls in COVID-19

## Identifying the type of design

Table 1: Summary definitions of policy impact evaluation designs commonly used for COVID-19

| Design | Units (e.g., regions of comparison) | | Time points measured per unit | | Assumed counterfactual. "If not for the intervention, ___" | Examples in the COVID-19 literature‡ |
|---|---|---|---|---|---|---|
| | With intervention | Without intervention | Before intervention | After intervention | | |
| Cross-sectional | At least one | At least one | N/A | One time point | Outcome in intervention units would have been the same as the outcome in the non-intervention units. | 14 |
| Pre/post Figure 1A | At least one | None | At least one (typically one) | At least one (typically one) | Outcome would have stayed the same from the pre period to the post period. | 15–22 |
| Interrupted time-series (ITS) Figure 1B | At least one | None | More than one | At least one (typically several) | Outcome slope and level* would have continued along the same modelled trajectory from the pre-period to the post period. | 23–36 |
| Difference-in-differences (DiD) Figure 1C | At least one | At least one† | At least one (typically one) | At least one (typically one) | Outcome in intervention units would have changed as much as (or in parallel with) the outcome in the non-intervention units. | 37–45 |
| Comparative interrupted time series (CITS) Figure 1D | At least one | At least one† | More than one (typically several) | At least one (typically several) | Outcome slope and level* would have changed as much as non-intervention group's slope and level* changed. | 46–49 |

* Assessing both slope and level only applicable if there are multiple data points during the post period
† Units without the intervention may be the pre-period of a different unit that eventually receives the intervention.
‡ As determined by reviewers in Haber et al;, 2021[50]

For our purposes, we define impact evaluation as examining the quantitative causal effect of particular policies (already implemented) on outcomes primarily using observational data. This is in contrast to models of hypothetical scenarios and/or using assumed or estimated data (e.g., typical uses of mechanistic or curve-fitting infectious disease models, scenario projections, etc.), noting that the lines between "models" and "impact evaluation" is blurry; models and impact evaluation methods are often together as complements within the same article.

COVID-19-related policy evaluation analyses typically fall under one of the categories in Table 1. In most cases, the design can be categorized using a combination of whether the analysis includes units (i.e. geographical regions, government groups, hospitals, etc) that did not receive the treatment (columns 2-3) and whether there are time points both before and after intervention for those units (columns 4-5). The penultimate column describes the implied counterfactual, discussed further in subsequent sections.

Cross sectional designs typically compare units with vs without the treatment at single time points. Pre/post studies typically compare outcomes before and after a policy, for units who received the intervention. Interrupted time-series analyses compare outcomes across time within units that received the intervention, and are used when there are at least two time points available before the intervention and at least one time point (typically multiple) after the intervention. ITS is distinct from pre/post in that it is based on data from multiple time points before the intervention in order to model a counterfactual trend and deviation from that trend over time, rather than just change in levels. Difference-in-differences analysis compares the outcome change in units that received the intervention with those that did not (or have not yet), with at least one point before and one after the intervention. In scenarios with multiple periods, that may involve a comparison with the pre-policy period of one region with the post-period of a different region, even though all regions eventually receive the intervention. Comparative interrupted time-series uses both aspects — a change over time and a comparison group — to compare the observed change in slopes for the intervention group with the change in slope for the comparison group.

Methods descriptions may not always provide a precise or reliable guide to which of the design approaches has been used. Some studies do not explicitly name these designs (or may classify them differently); and these are only a small fraction of designs and frameworks that are possible to use for policy evaluation.[7,9,51] Studies may have data at multiple time points but are effectively cross-sectional.[52] DiD, ITS, and CITS designs based on repeated cross-sectional data are sometimes described as "cross-sectional"[42,53] instead of longitudinal. The term "event study" is often used to refer to studies with a single unit and one change over time resembling ITS,[40,43,54] but may refer to other designs. Although ITS is often used to describe changes in one unit, it may also refer to settings in which many treated units adopt an intervention over time.[22,26,27,33,37,55] Studies will also frequently employ multiple designs,[40,56] while others use more complex methods of generating counterfactuals.[57] Definitions of these terms vary widely, and the definitions above should be considered as guidance only.

## Policy impact evaluation design foundations for COVID-19

The simplest design is the cross-sectional analysis, which compares COVID-19 outcomes between units of observation (e.g., cities) at a single calendar time or time since an event, typically post-intervention, often referred to as an ecological design. This design is unlikely to be appropriate for COVID-19-related policy evaluations, but provides a useful starting point for reasoning about different designs. Just as with comparisons of non-randomized medical treatments, the localities that adopt a particular policy likely differ substantially from those that don't on both observed and unobserved characteristics on a number of dimensions, including epidemic status and timing. Because both the breadth of important observed confounders typically exceeds the number of observable units and because of the large volume of critically important unobserved confounding, conditional exchangeability is unlikely to be met for COVID-19 policy evaluation.

Figure 1: Longitudinal designs overview

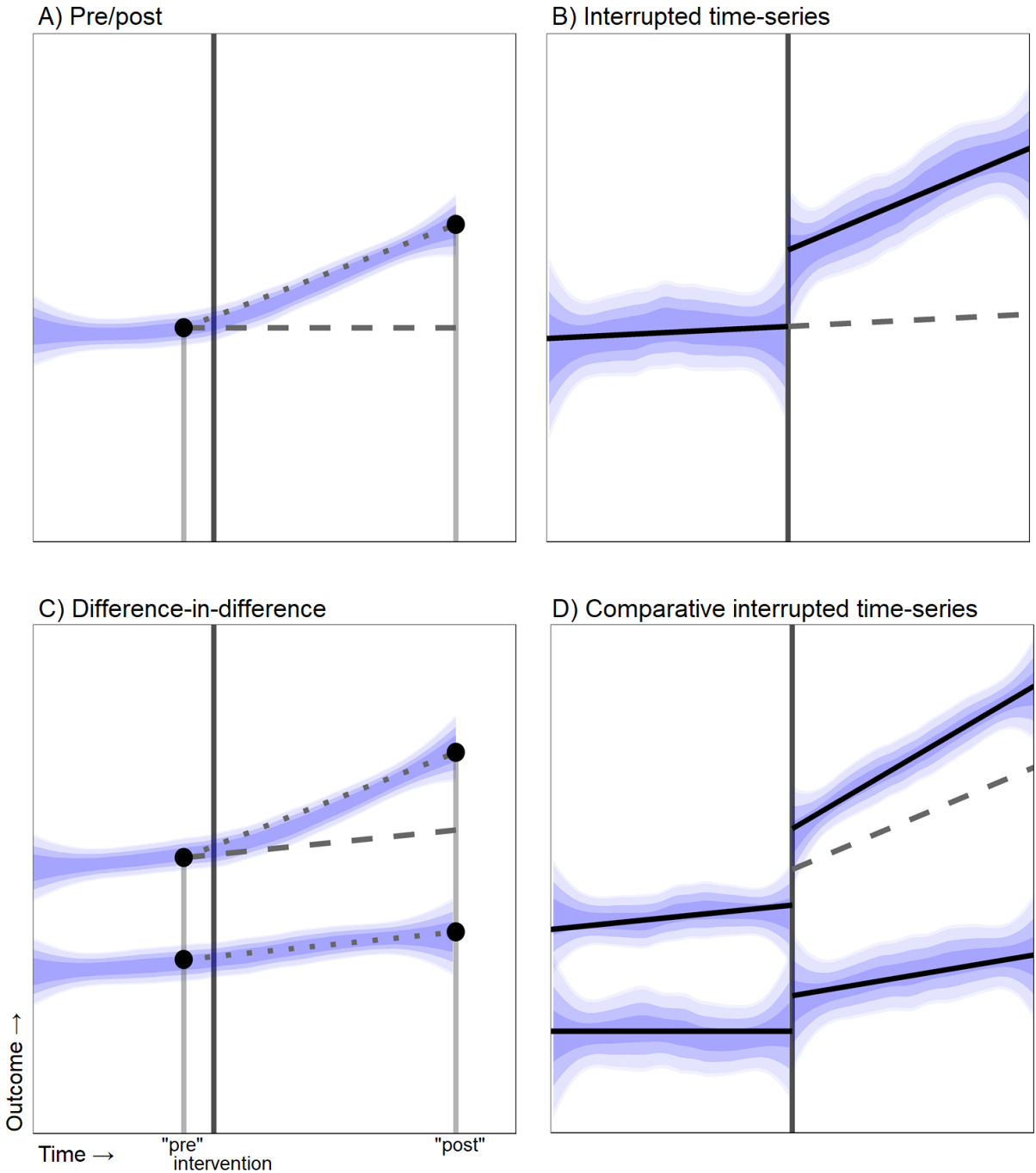

This chart shows four canonical longitudinal designs. In all cases: the blue shading represents the underlying data trends, the solid vertical grey line represents the time of intervention, the grey dashed lines represent the assumed counterfactual in the absence of the intervention, as discussed in the text. The impact estimate is obtained by comparing the outcomes observed for the treated unit in the post period (the solid line) with the implied counterfactual line (the dashed line). In the case of the pre/post and difference-in-differences panels the large black dots represent the time of measurement, connected by the grey dotted lines.

Given the limitations of simple cross-sectional comparisons, many evaluations of COVID-19 policies instead consider longitudinal designs, which look at differences or trends across time. These can be distinguished by the data used and the construction of the counterfactual, represented graphically for comparison in Figure 1 and summarized in text in Table 1. These methods construct a counterfactual of what would have happened in intervention regions based on past levels and trends in the outcomes, comparison with non-intervention comparators, or both.

## Pre/post studies

The simplest longitudinal design is a pre/post analysis, where some outcome is observed before policy implementation, and again after, in a single group (Figure 1A). Pre/post studies are analogous to a single arm trial[58] with no control and only a single follow-up observation after treatment. This effectively imposes the assumption that the counterfactual trend is completely flat, or that the outcome in the post-period in the absence of the policy change is the same as the value of the outcome before the policy change. Exchangeability is unlikely to hold between pre and post periods for COVID-19 policies, as this does not account for pre-existing underlying trends, and instead attributes all outcome changes completely to the intervention of interest. However, this is unlikely to be true in the case of COVID-19 policies. Just as the outcomes for an individual patient might be expected to change before and after treatment, for reasons unrelated to the treatment, outcomes related to policy interventions will change for reasons not caused by the policy. Infection rates, for example, would not be expected to remain stationary except in very specific circumstances, but a pre/post measurement would assume that any changes in infection rates are attributable to the policy.

## Interrupted time-series

Figure 2: Interrupted time-series graphical guidance for identifying common pitfalls

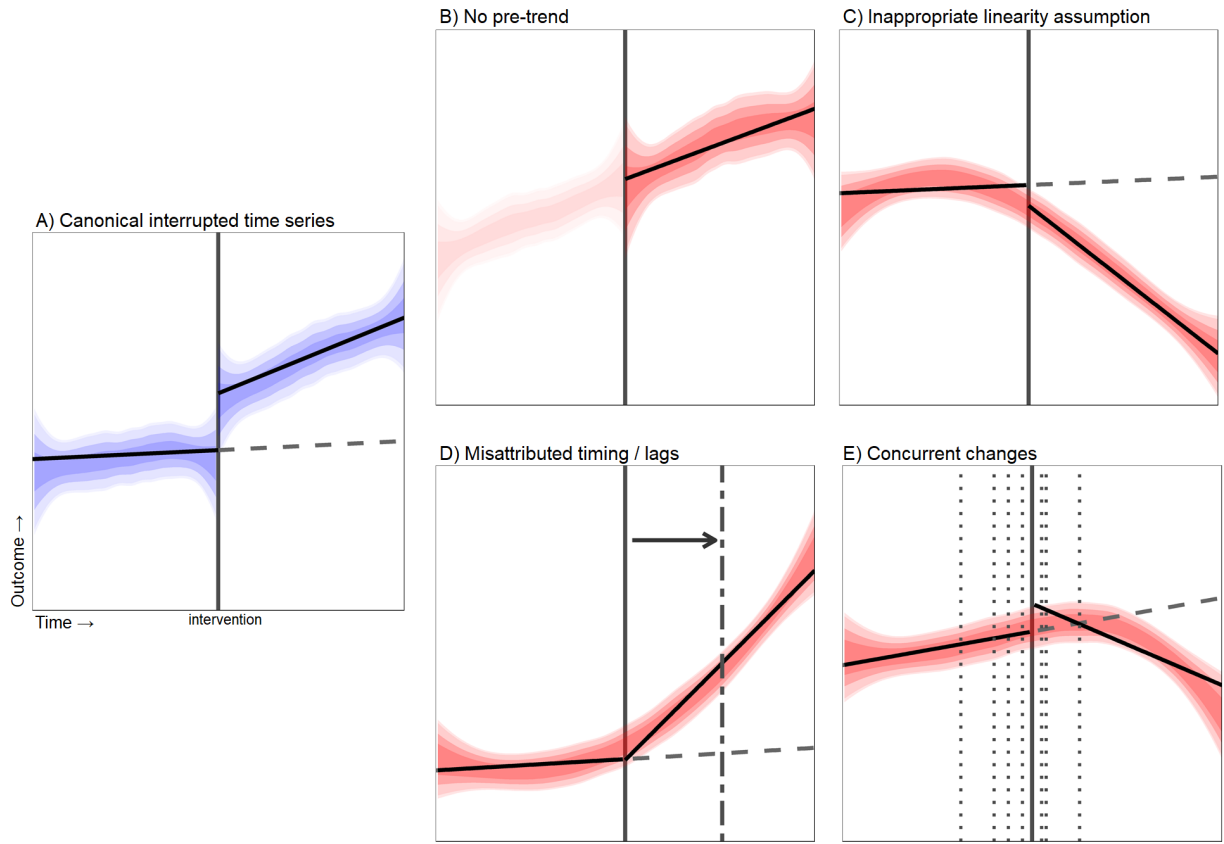

This chart shows one canonical design for ITS (blue, Panel A) and four panels demonstrating common issues with ITS analysis (red, panels B-E) discussed in the text. In all cases: the lag/red shading represents the underlying data trends, the vertical grey line represents the time of intervention, the grey dashed lines represent the assumed counterfactual in the absence of the intervention. In panel D) the dash-dot line represents the time at which the policy is expected to impact the outcome. In panel E), the vertical dotted lines represent concurrent events and changes.

Interrupted time-series (ITS) is a strategy that uses a projection of the pre-policy outcome trend as a counterfactual for how the outcome would have changed if the policy had not been introduced. In other words, in the absence of the policy change, ITS assumes the outcome would have continued on its pre-policy trend during the study period. ITS can be a useful tool in policy evaluation because it allows researchers to account for underlying trends in the outcome and, by comparing the treated unit (or location) to itself; it can therefore eliminate some of the confounding concerns that arise in cross-sectional or pre-post studies. Compared with a pre/post design, interrupted time-series requires a weaker exchangeability assumption because it allows and adjusts for prior trends.

However, the validity of ITS depends critically on how well counterfactual models in the outcome are correctly specified, and whether conditional exchangeability holds due to the policy of interest is the only relevant change during the study period. In the canonical setting (Figure 2A), the pre-policy trend is stable and can be modelled with the available data; the researcher appropriately models the timing of the change in the slope and/or level of the outcome; the researcher has sufficient information to conclude that there were no other changes during the study period that would be expected to influence the outcome. These elements are largely not satisfied in studies of COVID-related policy, as described below.

Visual and statistical examination of trends, preferentially alongside a theoretical justification of the model used, are key to examining assumptions for ITS. At a minimum, analyses should provide graphical representation of the data and model over time to examine whether pre-trend outcomes are stable, all trends are well-fit to the data, "interrupted" at the appropriate time point, and sensibly modelled (Figure 2B). In the case where an ITS includes a large number of units (e.g. states), it can be difficult to display this information graphically.

Model misspecification is a common pitfall in ITS (Figure 2C). The estimate of policy impact will be biased if a linear trend is assumed but the outcome and response to interventions instead follow nonlinear trends (either before or after the policy). In some cases, transformation of the outcome, for example using a log scale, may improve the suitability of a linear model, as in Palladino et al., 2020.[59] Imposing linearity inappropriately is a serious risk in the context of COVID-19, as trends in infectious disease dynamics are inherently non-linear.[54,60] For intuition, terms such as "exponential growth," "flattening," and "s-curves" all refer to non-linear infectious disease trends. Depending on the particular situation, non-linearity or other modelled trends can have complicated and counterintuitive impact on policy impact. Apparent linearity may also be temporary and an artifact of testing, which may give a misleading impression that linear models for infectious disease trends are appropriate indefinitely, as is the case for Zhang et al., 2020.[36] While some use linear progression in order to avoid more complex infectious disease models, in fact, linear projections impose strict and often unrealistic models, generally resulting in an inappropriate counterfactual. Functional form issues can often be mitigated through careful choice of models,[54] and/or testing of alternative assumptions.[61] At minimum, analyses should justify the use of their selected functional form, and preferably explore alternative specifications.

Researchers can easily misattribute the timing of the policy impact, resulting in spurious inference and bias (Figure 2D). Some public health policies can be expected to translate into immediate results (e.g., smoking bans and acute coronary events[54]). In contrast, nearly every outcome of interest in COVID-19 exhibits complex and difficult to infer time lags[62] typically in the realm of many weeks. The time between policy implementation and expected effect in the data can be large and highly variable. For example, in order to see the impact of a mask order, first the mask order takes effect, then people change their behaviors over time to comply with the order (or sometimes the reverse in the case of anticipation effects), mask use behavior produces changes in infections, then infections later result in symptoms, symptoms induce people to seek testing, the tests must then be processed in labs, and then finally the results get reported in data monitoring efforts. Selection of lead/lag time should be justifiable *a priori*, as was done in Islam et al., 2020[27] and Auger et al., 2020,[23] or external data. Selecting a lag based on the data, as in Slavova et al., 2020,[63] risks issues comparable to p-hacking,[64] where the lag is selected for the point at which results are more extreme, rather than most appropriate for causal inference.

Finally, and perhaps most concerningly in the context of COVID-19, conditional exchangeability fails for ITS when the policy of interest coincides in time with other changes that affect the outcome (Figure 2E).[61] COVID-19 policies rarely arrive alone; they are typically created alongside other policies, unofficial action, and large scale social and behavior changes[65] which themselves impact COVID-19-related outcomes. For example, if both mask and bar closure orders are rolled out together as a package, ITS cannot isolate the impact of bar closures specifically. In COVID-19 in particular, any number of factors can co-occur with policies, including (but not limited to) changing social behaviors, other policies, civic and political changes, and complicated infectious disease dynamics. These changes do not need to have taken place exactly concurrently with the policy implementation date of interest; they merely need to have some effect within the time period of measurement to result in potentially serious bias in effect estimates if unaddressed. Analogously, if an RCT involves randomizing people to a group receiving both A and B vs. control, we typically can't disentangle the effects of A from the effects of B, unless we also have separate A- and B-only arms. Ultimately, if multiple things are changing at the same time, ITS may not be an appropriate design for policy evaluation.

ITS will also likely be biased if, during the study period, there is a change in the way the outcome data is collected or measured. This might occur if the introduction of a COVID-19 control policy is combined with an effort to collect better data on infection or mortality cases.

Anticipation of a policy may induce behavior change before the actual policy takes effect. The policies themselves may have been chosen due to the expectation of change in disease outcomes, which introduces additional biases related to "reverse" causality.

Table 2: Checklist for identifying common pitfalls for ITS to evaluate COVID-19 policy

| **Key design questions.** If any answer is "no," this analysis is unlikely to be appropriate or useful for estimating the impact of the intervention of interest. | **Details and suggestions for identifying issues:** |
| --- | --- |
| Does the analysis provide graphical representation of the outcome over time? | -Check for a chart that shows the outcome over time, with the dates of interest. Outcomes may be aggregated for clarity (e.g. means and CIs at discrete time points). |
| Is there sufficient pre-intervention data to characterize pre-trends in the data? | -Check the chart(s) to see if there are several time points over a reasonable period of time over which to establish stability and curvature in the pre-trends. |
| Is the pre-trend stable? | -Check if there are sufficient data to reasonably determine a stable functional form for the pre-trends, and that they follow a modelable functional form. |
| Is the functional form of the counterfactual (e.g. linear) well-justified and appropriate? | -Check whether the authors explain and justify their choice of functional form. -Check if there is any curvature in the pre-trend. |

|  | -Consider the nature of the outcome. Is it sensible to measure the trend of this outcome on the scale and form used? Note: infectious disease dynamics are rarely linear.<br>-Consider that while pre-trend fit is a necessary condition for an appropriate linear counterfactual model, it is not sufficient. Check if the authors provide justification for the functional form to continue to be of the same functional form (e.g. linear). |
|---|---|
| Is the date or time threshold set to the appropriate date or time (e.g. is there lag between the intervention and outcome)? | -Check whether the authors justify the use of the date threshold relative to the date of the intervention.<br>-Trace the process between the intervention being put in place to when observable effects in the outcome might appear over time.<br>   -Consider whether there are anticipation effects (e.g. do people change behaviors before the date when the intervention begins?)<br>   -Consider whether there are lag effects. (e.g. does it take time for behaviors to change, behavior change to impact infections, infections to impact testing, and data to be collected, etc?)<br>   -Check if authors appropriately and directly account for these time effects. |
| Is this policy the only uncontrolled or unadjusted-for way in which the outcome could have changed during the measurement period? | -Consider any uncontrolled factor which could have influenced the outcome during the measurement period<br>-This may include (but is not limited to)<br>   -Other policies<br>   -Social behaviors<br>   -Economic conditions<br>   -Spillover effects from other regions<br>-Are these factors justified as having negligible impact?<br>   -If justified, is the argument that these have negligible impact convincing?<br>-Note that the actual concurrent changes do not need to happen during the period of measurement, just their effects. |

These issues are summarized as a checklist of questions to identify common pitfalls in Table 2.

## Difference-in-differences

The difference-in-difference (DiD) approach uses concurrent non-intervention groups as a counterfactual. Typically, this consists of one set of units (e.g., regions) that had the intervention and one set that did not, with each measured before and after the intervention took place. DiD is more directly analogous to a non-randomized medical study with at least one treatment and control group but limited observation before and after treatment. In contrast to ITS, which compares a unit with itself over time, DiD compares differences between treatment arms or units at two observation points. The exchangeability assumption in DiD applies to the outcome trends rather than the outcome levels; in other words, it implies that the treated and control units would have the same outcome *trend* in the absence of treatment. In many analyses, a DiD approach is implied by comparing regions over time, without formally naming or modelling it. Other DiD approaches use interventions implemented at multiple time points.[66] See Caniglia and Murray 2020[6] for a general overview of DiD in epidemiology and Goodman-Bacon and Marcus 2020[12] for further discussion of the nuances of using those methods to study COVID-19 related policies.

Figure 3: Difference-in-differences graphical guidance for identifying common pitfalls

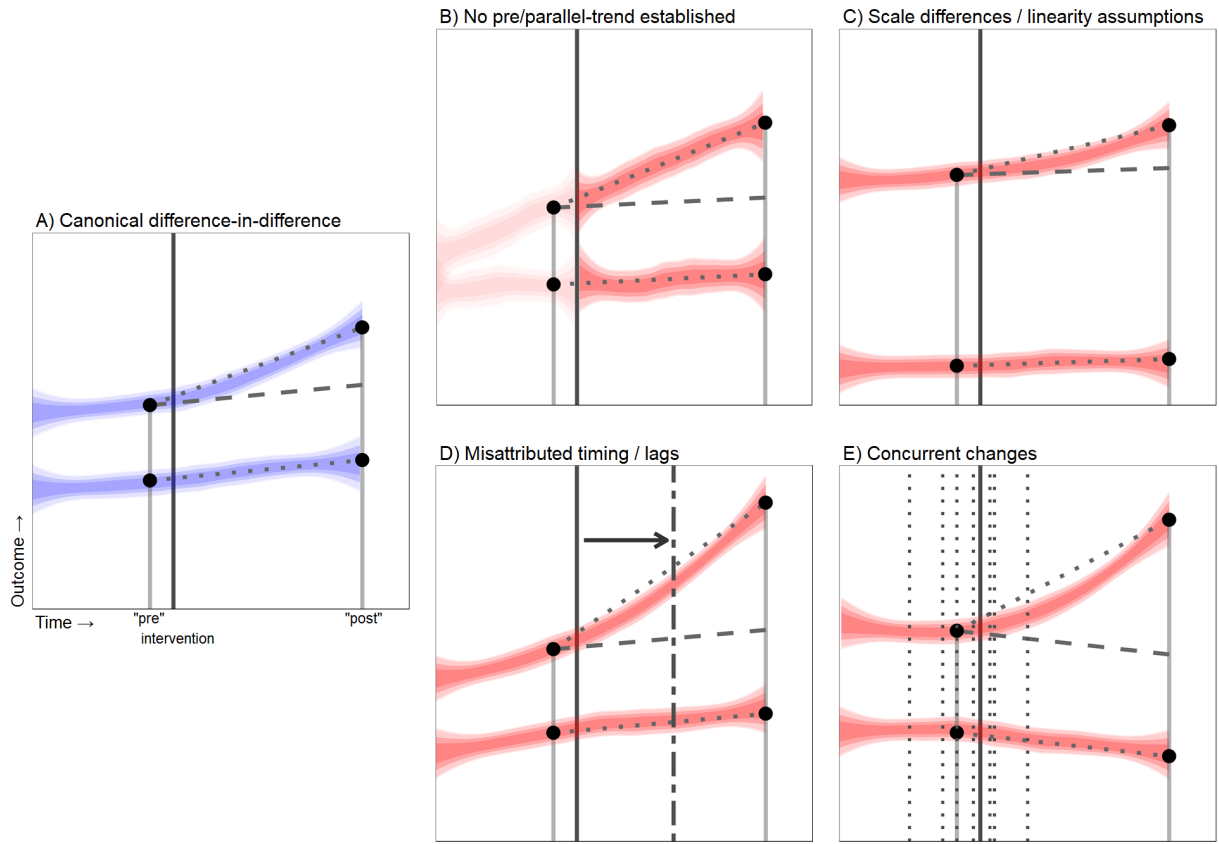

This chart shows one canonical design for DiD (blue, Panel A) and four panels demonstrating common issues with DiD analysis (red, panels B-E). In all cases: the blue/red shading represents the underlying data trends, the vertical grey line represents the time of intervention, the grey dashed lines represent the assumed counterfactual in the absence of the intervention. In panel D) the dash-dot line represents the time at which the policy is expected to impact the outcome. In panel E), the vertical dotted lines represent concurrent events and changes.

One key component of the standard DiD approach is the parallel counterfactual trends assumption: that the intervention and comparison groups would have had parallel trends over time in the absence of the intervention. In some cases, the parallel trends assumption may be referenced or examined implicitly but not named.[42]

Ideally, pre-intervention trends would be shown to be clearly identifiable, stable, of a similar level, and parallel between groups, such as in Hsaing et al 2020.[26] With only one observation before and only one after the intervention, assessment of the plausibility of the parallel counterfactual trends assumption is not possible. Absent this confirmation[67] the evaluation runs the risk of biased estimation due to differential pre-trends (Figure 3B). Pre-trends approaching the ceiling or floor[42] may also not be informative about stable and parallel pre-trends. Empirical assessment of whether pre-intervention trends were parallel and stable between groups is possible when multiple observations are available at multiple time points before the intervention, noting that this can begin to resemble a CITS design.[68] In this scenario, pre-trend data should be visually and statistically established and documented. While parallel trends before intervention (which we can observe and may be testable) do not guarantee parallel *counterfactual* trends in the post-intervention period (which we cannot observe and are generally untestable), examining pre-intervention parallel trends is a minimal requirement for DiD reliability.

Although DiD can provide some robustness for functional form, this can still be a major issue in many circumstances.[69] It is important to consider the scale and level on which the outcome is measured (Figure 3C) in order to correctly specify the outcome model. As with ITS, if the outcomes in the treatment and comparison groups are moving in parallel on a logged scale, they will not be moving in parallel on a natural scale. Level differences by themselves may be a problem for COVID-19 outcomes, as infectious disease transmission dynamics dictate that infection risks are related to the prevalence of infected people in a population, i.e. the rate of change is linked intrinsically to the level. A population with an extremely low prevalence will tend to have an inherently slower rise in infection rates than an otherwise identical population with merely a low prevalence. Just as importantly, large level differences in the outcome between intervention and comparison groups is often indicative of other important differences between comparators, which may result in other assumptions being violated.

While DiD is in some ways more robust to very specific kinds of timing effects (Figure 3D) and concurrent changes (Figure 3E), it also introduces additional risks. DiD effectively doubles the opportunity for concurrent changes to spuriously impact results, since they can occur in the treatment or comparison groups. As above, this can become even more problematic for DiD in the typical case where intervention groups enact more or very contextually different policies than non-intervention groups. If a concurrent change in the outcomes differs between the treated and comparison groups (e.g. one group implements the change and the other does not), then this would be considered a confounder of the DiD design. Even cases where concurrent changes happen equally in both treatment and comparison groups can lead to bias, particularly when approaching the maximum or minimum levels of the outcome. If either the treatment or control group is approaching the floor (e.g. 0% prevalence) or ceiling for an outcome of interest due to other policies concurrent in both places (e.g. national lockdowns, but region-level differences in mask policy), this can lead to bias when comparing changes between the two groups.

There are also risks of spillover effects (violations of the assumption of no interference). For example, infectious disease growth rates in one region can cause changes in growth rates in other regions, primarily through cross-border travel. Additionally, policies implemented in one region may influence behavior in a neighboring region, causing a change in the outcome trend in an untreated area.

Table 3: Checklist for identifying common pitfalls for DiD to evaluate COVID-19 policy

| Key design questions. If any answer is "no," this analysis is unlikely to be | Details and suggestions for inspection: |
| --- | --- |
| appropriate or useful for estimating the impact of the intervention of interest | |
| Does the analysis provide graphical representation of the outcome over time? | -Check for a graph that shows the outcome over time for all groups, with the dates of interest. Outcomes may be aggregated for clarity (e.g. mean and CI at discrete time points). |
| Is there sufficient pre-intervention data to observe both pre and post trends in the data? | -Check the chart(s) to see if there are several time points over a reasonable period of time over which to establish stability and curvature in the pre- and post- trends. |
| Are the pre-trends stable? | -Check if there are sufficient graphical data to reasonably determine a stable functional form for the pre-trends, and that they follow a modelable functional form. |
| Are the pre-trends parallel? | -Observe if the trends in the intervention and comparison groups appear to move together at the same rate at the same time. |
| Are the pre-trends at a similar level? | -Check if the trends in the intervention and comparison groups are at similar levels.<br>-Note that non-level trends exacerbates other problems with the analysis, including linearity assumptions |
| Are intervention and non-groups broadly comparable? | -Consider areas where comparison groups may be dissimilar for comparison beyond just the level of the outcome. |
| Is the date or time threshold set to the appropriate date or time (e.g. is there lag between the intervention and outcome)? | -Check whether the authors justify the use of the date threshold relative to the date of the intervention.<br>-Trace the process between the intervention being put in place to when observable effects in the outcome might appear over time.<br>    -Consider whether there are anticipation effects (e.g. do people change behaviors before the date when the intervention begins?)<br>    -Consider whether there are lag effects. (e.g. does it take time for behaviors to change, behavior change to impact infections, infections to impact testing, and data to be collected, etc?)<br>    -Check if authors appropriately and directly account for these time effects. |
| Is this policy the only uncontrolled or unadjusted-for way in which the outcome could have changed during the measurement period, differently for policy and non-policy regions? | -Consider any uncontrolled factor which could have influenced the outcome during the measurement period.<br>-Did any factor(s) influence the outcome different amounts in policy and non-policy regions?<br>-This may include (but is not limited to)<br>  -Other policies<br>  -Social behaviors<br>  -Economic conditions<br>  -Spillover effects from other regions<br>-Are these factors justified as having negligible impact?<br>  -If justified, is the argument that these have negligible impact convincing?<br>-Note that the actual concurrent changes do not need to happen during the period of measurement, just their effects. |

Similarly to the ITS section, these issues are summarized as a checklist of questions to identify common pitfalls in Table 3.

## Comparative interrupted time-series

Comparative interrupted time-series is conceptually and strongly related to DiD, but uses control locations to model the counterfactual change in trajectory over time, rather than just levels. The key difference lies in how pre-trends are modeled, which are not inherently assumed to be parallel in CITS. CITS allows controlling for differences in the overall trajectory, provided stable trends otherwise un-impacted by functional form, concurrent events, and other issues. In many cases, including the multiple-time-period case, DiD and CITs can be effectively indistinguishable.[68] As such, the same checks that apply to DiD also apply to CITS.

## Additional considerations

As with any other causal inference design, problems with measurement, generalizability, changes in measurement over time (e.g. varying test availability), statistical models, data quality, testing robustness to alternative assumptions, and many issues can undermine an otherwise robust evaluation, and are not discussed here. Decision-makers and researchers should pay particular attention to the relevance of the intervention as it was evaluated to relevant decisions being made. The impact of a program encouraging voluntary mask use may not tell us much about policies mandating the use of masks.

Policies are not selected at random, and additional consideration is needed for policy selection processes, especially when it concerns expected changes in the outcomes and/or other regions' policies impacting each other's policies. While the designs described here are mainly focused on addressing issues of confounding bias, bias can also arise from the way in which units are selected into the analysis. The treatment effect in the analyzed units may not be representative of the treatment effect in the population of interest.[70] More complex selection bias structures may also occur if, for example, selection into a DiD analysis is directly affected by (or is affected by variables associated with) both the treatment of interest and the outcome trend.[10]

# Discussion

In recent months, there has been a proliferation of research evaluating policies related to the COVID-19 pandemic. As with other areas of COVID-19 research, quality has been highly variable, with low quality studies resulting in poorly or mis-informed policy decisions, poorly tuned and inaccurate projection models, wasted resources, and undermined trust in research.[71,72] To support high quality policy evaluations, in this paper we describe common approaches to evaluating policies using observational data, and describe key issues that can arise in applying these approaches. This guidance should be considered minimal screening to identify low quality policy impact evaluation in COVID-19, but is in no way sufficient to identify high quality evidence or actionability. We hope that this guidance can help support researchers, editors, reviewers, and decision-makers in conducting high quality policy evaluations and in assessing the strength of the evidence that has already been published.

Training in epidemiologic methods for causal inference typically focuses on population-level inference from individual level data. Because policy evaluation is more often performed at the group level (e.g., geographic or government units), epidemiologists should be cautious when considering levels of analysis. As an analogy, the "individual" in policy inference is the government unit; and inference is being made for a "population" of government units. As such, the ecological fallacy—a fallacy generated when the exposure and outcome are at a lower level, but data are aggregated to a higher group level—does not necessarily apply; there is generally no conflict between the exposure (e.g. policy) and the outcome (e.g. the case counts within a government unit) and the level of inference. Hierarchical, clustered, and other multilevel designs are statistical models often used to improve statistical efficiency, examine mechanisms and secondary effects, and/or examine impacts in different subpopulations. However, for policy analysis, the statistical models are largely for estimation purposes, whereas the study design underlying the counterfactual construction at the policy level will still generally fall under the framework as specified in this guidance.

Policy evaluation — far from a simple task in normal circumstances — is particularly challenging during a pandemic, as highlighted in our recent systematic review of the COVID-19 policy literature based on the criteria in this guidance.[50] Cross-sectional comparisons of states or countries are likely to be biased by selection into treatment: for example, countries with worse outbreaks may be more likely to implement policies such as mask requirements. In analyses of changes over time – such as single-unit studies using interrupted time-series or multi-unit comparisons using difference-in-differences or comparative interrupted time-series – it may not be possible to parse apart the effects of different policies implemented around the same time, such as mask mandates paired with limits on social gatherings. Analyses of changes over time may also be biased if disease or human behavioral dynamics are not modelled appropriately. This can be challenging because case counts typically do not grow linearly and there is often a lag between a policy change and a behavioral response.

While this guidance is not comprehensive, it may help inform study designs not covered here. Many synthetic control methods,[73] and non-randomized trials with regional interventions (e.g. policies), are broadly comparable with the issues with difference-in-differences analyses we discuss here. Many causal inference methods for policy evaluation share much of their basic design and key assumptions with the designs in this guidance, although this might vary slightly depending on how they are structured. Other approaches may include adjustment and matching based observational causal inference designs,[10] instrumental variables and related quasi-experimental approaches,[7,9,74] cluster-randomized controlled trials. Each has its own set of practical, ethical, and inferential limitations.

In the face of these challenges, we recommend careful scrutiny and attention to potential sources of bias in COVID-19-related policy evaluations, but we remain optimistic about the potential for robust evaluations to inform decision-making. Researchers and decision-makers should triangulate across a large variety of approaches from theory to evidence, invest in better data and more reliable and useful evidence wherever feasible, clearly acknowledge limitations and potential sources of bias, and acknowledge when actionable evidence is not feasible.[75] We anticipate increasing opportunities for better examining policies moving forward, particularly if policies and interventions are designed with policy impact evaluation and data collection in mind.

The COVID-19 pandemic requires urgent decisions about policies that affect millions of people's lives in significant ways. High-quality evidence on the effects of these policies is critical to informing decision-making, but is difficult to generate, particularly under these extraordinary circumstances. Evidence-based decision-making and research depends on being able to quickly and efficiently synthesize and evaluate the strength of any evidence, and this is especially true in the context of a public health emergency. We hope that the guidance in this study can facilitate this process and improve the usefulness of policy evaluations by considering potential sources of bias, and how to communicate the underlying assumptions and sources of uncertainty.

## Acknowledgements

We would like to thank Dr. Sarah Wieten, Dr. Cathrine Axfors, Dr. Mario Malicki, and Dr. Mollie Wood for edits and suggestions on the manuscript. Critical feedback and support was provided by Dr. Steven Goodman and Dr. John Ioannidis. Dr. Kevin Hassett provided inspiration to use a cubic polynomial model for the infectious disease outcome curves in the figures.

## Funding

No funding was allocated specifically for this article.